\documentclass[prl,aps,twocolumn,superscriptaddress,showpacs]{revtex4}
\usepackage[dvips]{graphicx,color}\usepackage[dvips]{graphicx,color}
\def\(({\left(}
\def\)){\right)}
\def\[[{\left[}
\def\]]{\right]}

\begin{document}
\title{
Temperature Chaos and Bond Chaos in the Edwards-Anderson
%the Four-Dimensional $\pm$ 
Ising Spin Glasses : \\
Domain-Wall Free-Energy Measurements
}

\author{M. Sasaki}
\affiliation{
Department of Applied Physics, Tohoku University, Sendai, 980-8579, Japan}

\author{K. Hukushima}
\affiliation{
Department of Basic Science, University of Tokyo, Tokyo, 153-8902, Japan}

\author{H. Yoshino}
\affiliation{
Department of Earth and Space, Osaka University, Toyonaka, 560-0043, Japan}
%\affiliation{
%Laboratoire de Physique Th\'eorique et Hautes Energies, Jussieu, 
%5eme \'etage, Tour 24, 4 place Jussieu, 75252 Paris Cedex 05, France}
\affiliation{
Laboratoire de Physique Th\'eorique et Hautes Energies, Jussieu, 
75252 Paris Cedex 05, France}

\author{H. Takayama}
\affiliation{
Institute for Solid State Physics, University of Tokyo, 
Kashiwa-no-ha 5-1-5, Kashiwa, 277-8581, Japan}

\date{\today}

\begin{abstract}
Domain-wall free-energy $\delta F$, entropy 
$\delta S$, and the correlation function, $C_{\rm temp}$, of $\delta F$ are 
measured independently in the four-dimensional $\pm J$ Edwards-Anderson (EA) 
Ising spin glass. The stiffness exponent $\theta$, the fractal dimension 
of domain walls $d_{\rm s}$ and the chaos exponent $\zeta$
are extracted from the finite-size scaling analysis 
of $\delta F$, $\delta S$ and $C_{\rm temp}$ respectively
well inside the spin-glass phase. 
The three exponents are confirmed to satisfy the 
scaling relation $\zeta=d_{\rm s}/2-\theta$ derived by the droplet theory 
within our numerical accuracy.
%where the $T=0$ fixed point dominantly governs thermodynamic properties of the system. 
We also study bond chaos induced by random variation of bonds,
and find that the bond and temperature perturbations yield the 
universal chaos effects described by a common scaling function 
and the chaos exponent. 
These results strongly support the appropriateness 
of the droplet theory for the description of chaos effect 
in the EA Ising spin glasses.
%the droplet theory of the
% chaos effects in the EA Ising spin glasses.
\end{abstract}

\maketitle
In randomly frustrated systems such as spin glasses, 
directed polymer in random media (DPRM) and vortex glasses, 
the equilibrium ordered state could be completely reorganized 
by an infinitesimally small change in environment~\cite{McKayBerker82,
Kitatani86,BrayMoore87,Ritort94,HuseKo97,Nifle98,SalesYoshino02b}. 
This curious property called chaos effect has attracted much attention
since it was found in 1980s~\cite{McKayBerker82}. 
Especially, chaos induced by temperature variation (temperature chaos) 
is now of great interest because of its potential relevance for rejuvenation 
caused by temperature variation~\cite{Rejuvenation}. 
However, the issue of temperature chaos still remains far from being resolved. 
In particular, concerning low-dimensional Edwards-Anderson (EA) 
Ising spin glass models, the situation is very controversial because 
numerical studies so far done provide the evidence both for and against 
temperature chaos~\cite{Nifle98,HuseKo97,BilloireMarinari00,
HukushimaIba02,AspelmeierBray02}. 

\begin{figure}[b]
\includegraphics[angle=0,width=7.5cm]{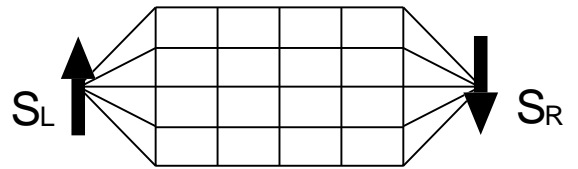} 
\caption{Model for the boundary flip MC.}
\label{fig:model}
\end{figure}

In the present work, we examine temperature chaos by numerical
measurements of the domain wall free-energy $\delta F$, the difference
in the free-energy between the system with the periodic boundary
condition (BC) and that with the anti-periodic BC.  This 
$\delta F$ relates to the effective coupling $J_{\rm eff}$ between 
the two boundary spins $S_{\rm L}$ and $S_{\rm R}$ (see Fig.~\ref{fig:model}) 
as $J_{\rm eff}=-\delta F/2$~\cite{BrayMoore84,McMillan85}. 
We find $\delta F$ of each sample exhibits oscillations
along the temperature axis providing direct evidence of the temperature  
chaos. 
Furthermore, we find from simultaneous observations of the 
domain wall energy $\delta E$ and so entropy $\delta S$ 
that $\delta E$ and $T \delta S$ are large but they cancel with each other 
in the leading order to yield significantly small 
$\delta F = \delta E - T\delta S$. 
Such intriguing behavior is indeed predicted by the droplet 
theory~\cite{DropletTheory}.
For a quantitative check of the droplet theory we focus on
the anticipated scaling relation 
\begin{equation}
\zeta=d_{\rm s}/2-\theta
\label{eqn:scaling-rela}
\end{equation}
derived from it, where the stiffness exponent $\theta$ is extracted 
from $\sigma_F \sim L^\theta$, the fractal dimension of domain walls 
$d_{\rm s}$ from $\sigma_{\rm S} \sim L^{d_{\rm s}/2}$, 
and the so called chaos exponent 
$\zeta$ from $C_{\rm temp}(L,T,T+\Delta) \sim f(L(\Delta T)^{-1/\zeta})$. Here 
$\sigma_F$ and $\sigma_S$ are the standard deviations of $\delta F$ and
$\delta S$, respectively, and $C_{\rm temp}$ is the correlation function
of $\delta F$'s defined by Eq.(\ref{eqn:CFtempDef}) below and $f(x)$ is
a certain scaling function. 
We find the three fundamental exponents thus 
extracted indeed satisfy Eq.~(\ref{eqn:scaling-rela}) well
inside the spin-glass phase whose thermodynamic properties are dominantly
governed by the $T=0$ fixed point. 

We also study bond chaos by measuring how $\delta F$ varies with changes
in couplings. The result evidently shows the existence of bond
chaos. Moreover, the scaling analysis of two correlation functions
associated with temperature and bond perturbations reveals
quantitatively that not only the chaos exponent but also the scaling
function are common to both the perturbations. This universal aspect of
chaos effect anticipated from the droplet theory~\cite{DropletTheory} is
also observed in the Migdal-Kadanoff spin 
glasses~\cite{BanavarBray87,NifleHilhorst92}
and the DPRM~\cite{SalesYoshino02b}. All of our numerical
results, particularly the quantitative check of 
Eq.~(\ref{eqn:scaling-rela}), are strong evidence not
only for the existence of chaos in the EA Ising spin glasses but also
for the appropriateness of the droplet theory for its description.
%{\color{red}
%The difference between the two perturbations 
%is absorbed into the numerical prefactor 
%of the so called overlap length. As shown in eq.~\ref{eqn:Ell} below, 
%the prefactor is relatively larger for the temperature chaos case. 
%This explains why it has been 
%rather difficult to observe the temperature chaos effect numerically~\cite{Nifle98,AspelmeierBray02}}

{\it The boundary flip MC method---}
Let us first describe the boundary flip MC 
method~\cite{Hasenbusch93,Hukushima99} which enables us to measure the
domain-wall free-energy.  
We consider a model which consists of Ising spins on a 
$d$-dimensional hyper-cubic lattice of $L^d$ and two {\it boundary}
Ising spins $S_{\rm L}$ and $S_{\rm R}$ (see Fig.~\ref{fig:model}). 
The usual periodic BC is applied for the directions along which 
the two boundary spins do not lie. 
The Hamiltonian is 
${\cal H}=-\sum_{\langle ij \rangle} J_{ij} S_i S_j$, where the sum is
over all the nearest neighboring pairs including those consisting of one
of the two boundary spins and a spin on the surfaces of the lattice. 
In our boundary flip MC simulation, the two boundary spins 
are also updated according to a standard MC procedure. 
For each spin configuration simulated, we regard the BC 
as periodic (anti-periodic) when $S_{\rm L}$ and $S_{\rm R}$ are 
in parallel (anti-parallel). 
Since the probability $P_{\rm P(AP)}(T)$ for finding the periodic 
(anti-periodic) BC is proportional to $\exp[-F_{\rm P(AP)}(T)/T]$, where
$F_{\rm P(AP)}(T)$ is the free-energy with the periodic (anti-periodic)
BC, we obtain, with $P_{\rm AP}=1-P_{\rm P}$,
\begin{eqnarray}
\delta F(T)&\equiv& F_{\rm P}(T)- F_{\rm AP}(T) \nonumber \\
&=&-k_{\rm B}T\{\log[P_{\rm P}(T)]-\log[1-P_{\rm P}(T)]\}.
\end{eqnarray}
%It should be noted that $\delta F$ and the {\it effective} coupling 
%$J_{\rm eff}$ between $S_{\rm L}$ and $S_{\rm R}$ are related by 
%$J_{\rm eff}=-\delta F/2$~\cite{BrayMoore84,McMillan85}. 
We also measure the thermally averaged energy $E_{\rm P(AP)}(T)$ 
when the two boundary spins are in parallel (anti-parallel). 
It enables us to estimate the domain-wall {\it energy} 
$\delta E(T)\equiv E_{\rm P}(T)-E_{\rm AP}(T)$. 
Then, the domain-wall {\it entropy} $\delta S$ is evaluated
either from $\delta S=(\delta E -\delta F)/T$ or 
$\delta S=-\frac{\partial (\delta F)}{\partial T}$. 
We have checked that both the estimations yield identical
results within our numerical accuracy. 

We study the four-dimensional $\pm J$ Ising spin glasses in the present
work. In four dimensions  
%The reason why we study four dimensional spin glasses is that 
the value of the stiffness exponent $\theta$ is significantly 
large~\cite{Hukushima99,Hartmann99d}, which enables us to make scaling
analyses rather easily as compared in three dimensions.
The values of $\{J_{ij}\}$ are taken from a bimodal distribution with equal weights at $J_{ij}=\pm J$. 
We use the exchange MC method~\cite{HukushimaNemoto96} to accelerate the
equilibration.  
The temperature range we investigate is between $0.6J$ and $4.5J$, 
whereas the critical temperature of the model is 
around $2.0J$~\cite{MarinariZuliani99}. 
The sizes we study are $L=4,~5,~6,~7,~8$ and $10$. 
The number of samples is $824$ for $L=10$ and $1500$ for the others. 
The period for thermalization and that for measurement are set 
sufficiently (at least 5 times) larger than the ergodic time, 
which is defined by the average MC step for a specific 
replica to move from the lowest to the highest temperature and return 
to the lowest one. 
%We have confirmed the thermalization 
%by checking that the two boundary spins, which are the most difficult 
%spins to relax, forget their initial configuration before the measurement.

\begin{figure}[b]
\includegraphics[width=\columnwidth]{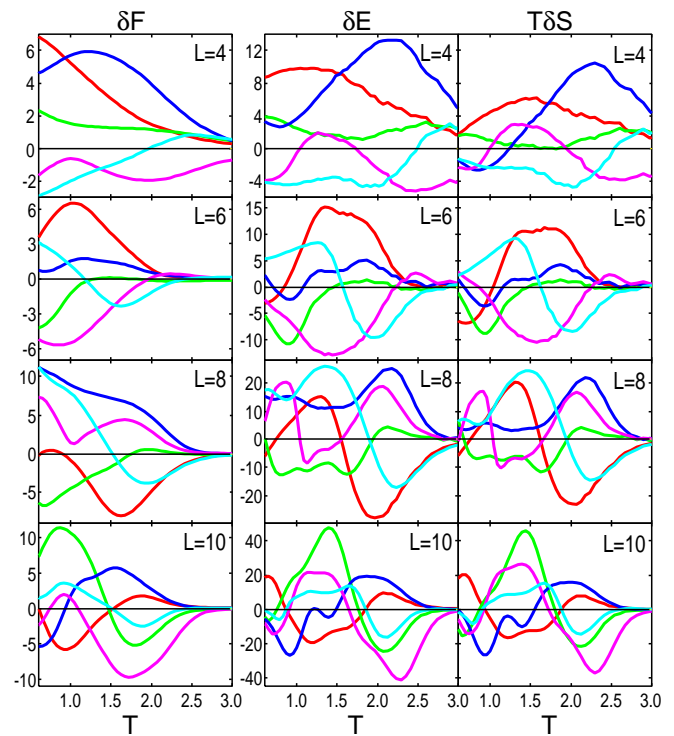} 
\caption{$\delta F$ (left), $\delta E$ (middle) and $T\delta S$ (right) vs. 
temperature for 5 samples. $L=4,~6,~8$~and~$10$ from top 
to bottom. 
}
\label{fig:5samples}
\end{figure}

{\it Temperature chaos---}
In Fig.~\ref{fig:5samples}, 
we show temperature dependence of $\delta F$, $\delta E$ and $T\delta S$ 
for 5 samples. 
Oscillations of the three observables become stronger with increasing $L$. 
We in fact see that $\delta F$ of some samples changes its sign, 
meaning that the favorable BC with the lower free-energy
changes with temperature. We also see that, as predicted 
by the droplet theory~\cite{DropletTheory}, 
$\delta E(T)$ and $T\delta S(T)$ exhibit very similar
temperature dependence and cancel with each other in the leading order 
to yield relatively small $\delta F$.
%Figure~\ref{fig:5samples} also shows that the amplitudes 
%of the three observables increase with $L$.

\begin{figure}[b]
\includegraphics[angle=270,width=\columnwidth]{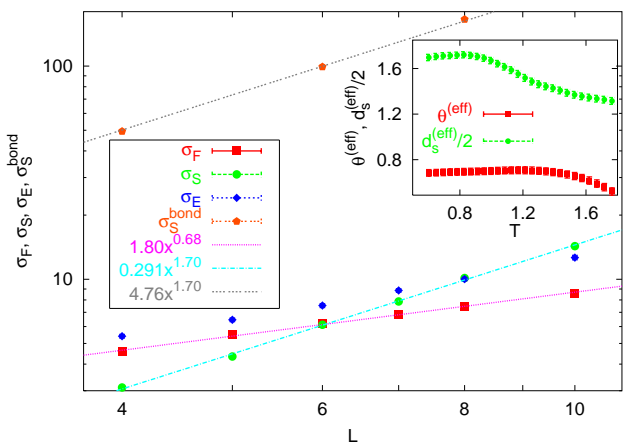}
\caption{Size dependences of $\sigma_{\rm F}$, $\sigma_{\rm E}$, 
$\sigma_{\rm S}$ and $\sigma_{\rm S}^{\rm bond}$ at $T=0.6J$. 
See text for their definitions. The two straight lines for $\sigma_{\rm F}$ 
and $\sigma_{\rm S}$ are obtained by linear least-square fits of 
$\ln(\sigma_{\rm F})$/$\ln(\sigma_{\rm S})$ against $\ln(L)$. 
The line for $\sigma_{\rm S}^{\rm bond}$ 
has the same slope as that for $\sigma_{\rm S}$. The inset shows the data 
for $\theta^{\rm (eff)}$ and $d_{\rm s}^{\rm (eff)}/2$. 
}
\label{fig:ThetaDs}
\end{figure}

In Fig.~\ref{fig:ThetaDs}, the standard deviations,  
%$\delta F$, $\delta E$ and $\delta S$, denoted respectively as 
$\sigma_{\rm F}$, $\sigma_{\rm E}$ and $\sigma_{\rm S}$, at $T=0.6J$ are
plotted as a function of $L$. Interestingly, $\sigma_{\rm S}$, which gives 
the amplitude of $|\frac{\partial (\delta F)}{\partial T}|$, 
increases more rapidly than $\sigma_{\rm F}$, i.e., the amplitude of
$\delta F$. See~\cite{AspelmeierBray02} for a similar observation 
in the three-dimensional EA model.
As argued by Banavar and Bray~\cite{BanavarBray87}, this result
naturally leads us to the conclusion that $\delta F$ 
in the limit $L\rightarrow \infty$ is totally temperature chaotic. 

%According to the droplet theory, at low enough temperature and large
%enough size, $\sigma_{\rm F}$ and $\sigma_{\rm S}$ are proportional to 
%$L^{\theta}$ and $L^{d_{\rm s}/2}$, respectively.
The inset of Fig.~\ref{fig:ThetaDs} shows $\theta(T)$ and 
$d_{\rm s}(T)/2$ estimated by linear least-square fits of
$\ln(\sigma_{\rm F})$ and $\ln(\sigma_{\rm S})$ against $\ln(L)$ at each
temperature. As expected from the droplet theory which is constructed
around the $T=0$ fixed point, the two exponents converge to a certain 
value at low temperatures. By averaging over the lowest five
temperatures, we obtain 
\begin{equation}
\theta = 0.69\pm 0.03,\hspace{5mm} d_{\rm s} = 3.42\pm 0.06.
\label{eqn:EthetaEds}
\end{equation}
Our $\theta$ is compatible with other 
estimations~\cite{Hartmann99d,Hukushima99}, while our $d_{\rm s}$ is
somewhat smaller than other ones~\cite{DSestimation}. 
The apparent temperature dependence of $\theta(T)$ and 
$d_{\rm s}(T)$ at higher temperatures is considered to be due to the
critical fluctuation associated with the unstable fixed
point at $T_{\rm c}$, combined with the finite-size effect. 
Its detailed quantitative analysis is, however,  
beyond the scope of the present work.

\begin{figure}[b]
\includegraphics[angle=270,width=\columnwidth]{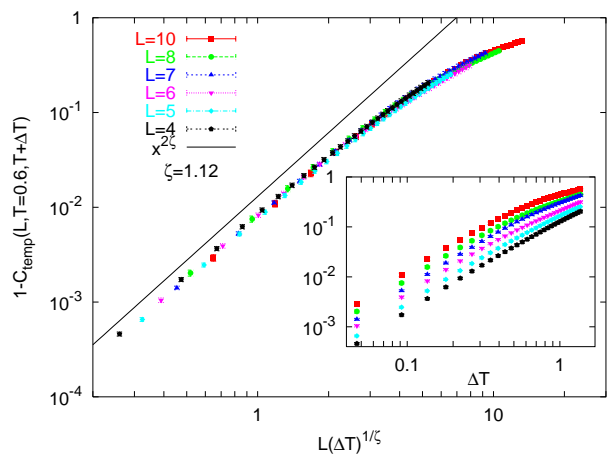} 
\caption{A scaling plot of $1-C_{\rm temp}(L,T,T+\Delta T)$ at $T=0.6J$ 
against $L(\Delta T)^{1/\zeta}$ 
with $\zeta=1.12$. 
The line is proportional to $x^{2\zeta}$. 
In the inset, $1-C_{\rm temp}$ for $L=4,~5,~6,~7,~8$~and $10$ 
(from bottom to top) are plotted as a function of $\Delta T$. }
\label{fig:CFtemp}
\end{figure}

We next examine the correlation function defined by 
\begin{equation}
C_{\rm temp}(L,T,T+\Delta T)\equiv \frac{\overline{\delta F(L,T)\delta F(L,T+\Delta T)}}
{\sigma_{\rm F} (L,T)\sigma_{\rm F} (L,T+\Delta T)},
\label{eqn:CFtempDef}
\end{equation}
where $\overline{\cdots}$ is the sample average.
%and $\sigma_{\rm F}(L,T)$ is the standard deviation of $\delta F(L,T)$. 
A similar correlation function was first introduced by Bray and Moore to
study bond chaos~\cite{BrayMoore87}. The inset of Fig.~\ref{fig:CFtemp}
shows the raw data of $1-C_{\rm temp}$ at $T=0.6J$. 
$C_{\rm temp}$ approaches zero rapidly with increasing $L$. 
From the prediction of the overlap length by the
droplet theory, over which the configurations at the two temperatures
are unrelated, we expect one parameter scaling of 
$C_{\rm temp}=f(L\Delta T^{1/\zeta})$ whose test is shown in the main
frame of  Fig.~\ref{fig:CFtemp}. 
We see that the scaling works nicely. 
%The scaling indicates that an infinitesimal change in temperature completely destroys correlation of $\delta F$ in the limit $L\rightarrow \infty$. 
The value of $\zeta$ is evaluated to be $1.12 \pm 0.05$ by the fitting. 
Quite interestingly, this value is consistent with the value 
$\zeta=1.02 \pm 0.06$ obtained by substituting eq.~(\ref{eqn:EthetaEds}) 
into Eq.~(\ref{eqn:scaling-rela}) predicted by the droplet theory. 
This is one of the main results of the present work. 
We also see that the data are consistent with 
the expected asymptotic behavior in the limit 
$L^{\zeta}\Delta T \rightarrow 0$, 
$1-C_{\rm temp}\propto (L^{\zeta}\Delta T)^2$~\cite{BrayMoore87}, 
as depicted by the line. 
%From these results, we conclude that temperature chaos does exist in the EA spin glasses of the present interest. 

%Now let us consider how the data are scaled. 
%Decorrelation of $\delta F(T)$ and $\delta F(T+\Delta T)$ begins 
%when the difference between the two 
%($\approx -\delta S(T)\Delta T$) is comparable with $\delta F(T)$ 
%itself. 
%Then, by recalling that $\sigma_{\rm F}\propto L^{\theta}$ and 
%$\sigma_{\rm S}\propto L^{d_{\rm s}/2}$, the overlap length 
%$\ell_{\rm temp}$, where the decorrelation starts, is considered to be
%proportional to $\Delta T^{-1/\zeta}$, where 
%$\zeta\equiv d_{\rm s}/2-\theta$ is the chaos exponent 
%introduced in the droplet theory. 
%This argument suggests that 
%$L \Delta T^{1/\zeta}(\propto L/\ell_{\rm temp})$ 
%is a proper scaling variable. The main frame of  Fig.~\ref{fig:CFtemp} 
%is the result of the scaling plot. 
%All the data in the inset nicely merge into a single curve. 

{\it Bond chaos and universality---}
We also study bond chaos by comparing two systems with correlated coupling 
sets. The perturbed couplings $\{ J_{ij}' \}$ are obtained from the
unperturbed ones $\{ J_{ij} \}$ by changing the sign of $J_{ij}$ with
probability $p$. Since simulation for bond chaos 
costs much more time than that for temperature chaos, 
we only examined $L=4,6,8$ for bond chaos. 

\begin{figure}[t]
\includegraphics[angle=270,width=\columnwidth]{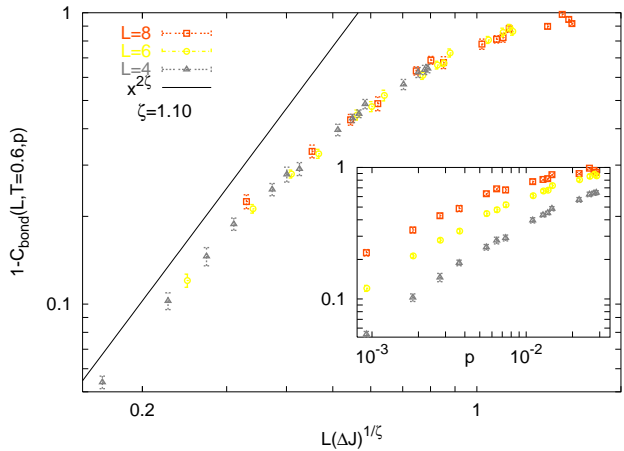} 
\caption{A scaling plot of $1-C_{\rm bond}(L,T,p)$ at $T=0.6J$ 
against $L(\Delta J)^{1/\zeta}$ 
with $\zeta=1.10$, where $\Delta J = \sqrt{p}$. 
The line is proportional to $x^{2\zeta}$. 
In the inset, the raw data for $L=4,~6$~and $8$ (from bottom to top) 
are shown as a function of $p$.
}
\label{fig:CFbond}
\end{figure}

Now let us consider an observable 
$\delta S^{\rm bond}\equiv -\frac{\delta F'-\delta F}{\Delta J}$, 
where $\Delta J\equiv \sqrt{p}$ and $\delta F$ ($\delta F'$) is the
domain-wall free-energy of the unperturbed (perturbed) system. 
$\delta S^{\rm bond}$ here and $\delta S$ discussed above are similar in
a sense that the both are the increment ratios of $\delta F$ against the
perturbations.  The ratio against $\Delta J$, not $p$
itself, is considered to compare temperature perturbation and bond
perturbation properly~\cite{Nifle98}. In Fig.~\ref{fig:ThetaDs}, the
standard deviation of $\delta S^{\rm bond}$, denoted as 
$\sigma_{\rm S}^{\rm bond}$, is also shown.  
$\delta S^{\rm bond}$ is estimated with $\Delta J\approx 0.03$, which
corresponds to a small value of $p\approx 0.0009$. 
The line for $\sigma_{\rm S}^{\rm bond}$ and that for $\sigma_{\rm S}$ 
have the same slope, which suggests that temperature and
bond perturbations belong to the same universality class. 
The coefficient of $\sigma_{\rm S}^{\rm bond}$ is, however, about $16.4$
times as large as that of $\sigma_{\rm S}$.  
In the inset of Fig.~\ref{fig:CFbond} we show the raw data of the
correlation function for bond perturbation defined by 
\begin{equation}
 C_{\rm bond}(L,T,p)\equiv \frac{\overline {\delta F(L,T) \delta F'(L,T)}}
{\sigma_{\rm F} (L,T)\sigma_{\rm F}'(L,T)}. 
\end{equation}
Again, the correlation decays faster with increasing $L$. 
In the main frame of Fig.~\ref{fig:CFbond}, we test a similar 
scaling to that in Fig.~\ref{fig:CFtemp} by assuming that the overlap length 
of the bond perturbation scales as $\Delta J^{-1/\zeta}$. 
All the data again collapse into a single curve.
The chaos exponent $\zeta$ is evaluated to be 
$1.10 \pm 0.10$ by the fitting. 
%which agrees with the estimation by Nifle~\cite{Nifle98}. 
%Our results indicate that the chaos exponents for both temperature and
%bond perturbations are the same in the spin glass phase 
%(below $T_{\rm c}$), as is the case in the critical
%region (at $T_{\rm c}$)~\cite{Nifle98}.  

\begin{figure}[t]
\includegraphics[angle=270,width=\columnwidth]{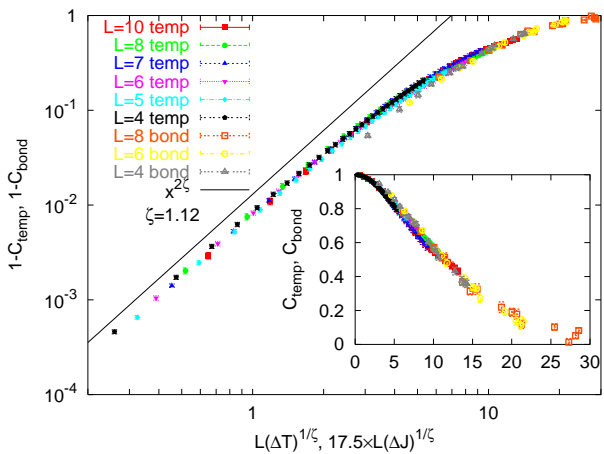} 
\caption{A scaling plot of $1-C$ by using all the data 
in Figs.~\ref{fig:CFtemp}~and~\ref{fig:CFbond}. 
The scaling variable for the data of temperature perturbation is 
$L(\Delta T)^{1/\zeta}$, while that for the data of bond perturbation 
is $17.5\times L(\Delta J)^{1/\zeta}$, where $\zeta=1.12$.
The line is proportional to $x^{2\zeta}$. 
The inset shows the same plot for $C$.}
\label{fig:CFScalingA}
\end{figure}

To compare the two scaling functions for temperature and bond perturbations, 
we plot in Fig.~\ref{fig:CFScalingA} all the data of both $C_{\rm temp}$ and 
$C_{\rm bond}$ by using the same chaos exponent. 
Here we use the value $\zeta=1.12$ in
Fig.~\ref{fig:CFtemp}, and multiply the scaling variable $L(\Delta
J)^{1/\zeta}$ by factor~17.5. All the data roughly merge into a single
curve, indicating that the chaos exponent and the scaling function for
temperature chaos are the same as those for bond chaos. 
%The factor 17.5 again shows that bond perturbation is much 
%stronger than bond perturbation.
%The factor 17.5 is roughly consistent with an expected value 
%$(\sigma_{\rm S}^{\rm bond}/\sigma_{\rm S})^{1/\zeta}=15.5$, 
%where we have used the value $\zeta=1.02$ coming 
%from eq.~(\ref{eqn:EthetaEds}).
Lastly, by estimating the overlap length $\ell$ as the value of 
$L$ for which $C=0.5$, we obtain
\begin{equation}
\ell_{\rm temp}\approx 11.5\Delta T^{-1/\zeta},\hspace{5mm}\ell_{\rm bond}
\approx 0.657\Delta J^{-1/\zeta},
\label{eqn:Ell}
\end{equation}
where $\zeta\approx 1.12$. We see that the overlap length of 
the bond perturbation is much shorter than that of the temperature one.
This result as well as factor 16.4 in the 
$\sigma_{\rm S}^{\rm bond}$ scaling mentioned above are the quantitative
description of the well-known fact that temperature chaos is much more
difficult to be observed than bond chaos~\cite{Nifle98,AspelmeierBray02}.

{\it Conclusion---}
In the present work, we have studied the four-dimensional EA 
Ising spin glass with focus on the chaos effect.
As a consequence, many non-trivial predictions by the
droplet theory, such as the $\delta F$ oscillation along the temperature
axis and the cancellation of $\delta E$ and $T\delta S$ are found in this
model. Most importantly, the scaling relation of
Eq.~(\ref{eqn:scaling-rela}) and the universal aspect of temperature and
bond chaos effects are quantitatively confirmed well inside its
spin-glass phase whose thermodynamic properties 
are dominantly governed by the $T=0$ fixed point. 
These results are certainly strong evidences
for the appropriateness of the droplet theory for the description of 
chaos effect in the EA Ising spin glasses.
On the other hand, recent work by Rizzo and 
Crisanti~\cite{RizzoCrisanti02} indicates the existence of similar 
chaos effects in the Sherrington-Kirkpatrick model. 
Whether our results are consistent with the mean field view point or not
is an interesting open problem.

%Although we consider that these results 
%give some credit for the droplet theory, it is premature to believe 
%that the droplet theory is only the way to explain chaos in spin glasses. 
%In fact, temperature chaos may exist in the Sherrington-Kirkpatrick 
%model~\cite{RizzoCrisanti02}. It is quite interesting to investigate 
%whether our results are consistent with the mean-field viewpoint or not. 
%}

We would like to thank Dr. Katzgraber for fruitful discussion and useful 
suggestions. M.S. acknowledges financial support from the Japan Society 
for the Promotion of Science. The present work is supported by 
Grant-in-Aid for Scientific Research Program (\# 14540351, \# 14084204, 
\# 14740233) and NAREGI Nanoscience Project from the MEXT. 
The present simulations have been performed on 
SGI 2800/384 at the Supercomputer Center, Institute for Solid State Physics, 
University of Tokyo.

\end{document}